\documentclass[]{spie} 
\usepackage{latexsym}
\usepackage{times,mathptm} 
\usepackage{graphicx} 
\usepackage{multicol}
\usepackage{rotating} 
\usepackage[verbose]{wrapfig}
\usepackage{epsfig}
\newcommand{\g}{$\gamma$}

\newcommand{\cs}{$^{137}$Cs}

\newcommand{\na}{$^{22}$Na}
\newcommand{\co}{$^{60}$Co}
\newcommand{\yt}{$^{88}$Y}

\newcommand{\bc}{\begin{center}}
\newcommand{\ec}{\end{center}}
\newcommand{\be}{\begin{equation}}
\newcommand{\ee}{\end{equation}}
\newcommand{\bfg}{\begin{figure}}
\newcommand{\efg}{\end{figure}}
\newcommand{\bi}{\begin{itemize}}
\newcommand{\ei}{\end{itemize}}
\newcommand{\bt}{\begin{table}}
\newcommand{\enta}{\end{table}}
\newcommand{\keV}{\mbox{ke\hspace{-0.1em}V}}
\newcommand{\MeV}{\mbox{Me\hspace{-0.1em}V}}

\newcommand{\phibar}{\ensuremath{\bar{\varphi}}}

\newcommand{\phigeo}{\ensuremath{\varphi_\mathrm{geo}}}
\renewcommand{\deg}{\ensuremath{^\circ}}

\title{The LXeGRIT Compton Telescope Prototype: Current Status and Future
Prospects} 

\author{E.~Aprile\supit{a}, A.~Curioni\supit{a}, K.L.~Giboni\supit{a} ,
	M.~Kobayashi\supit{a}, \\ 
        U.G.~Oberlack\supit{b}, E.L.~Chupp\supit{c}, P.P.~Dunphy\supit{c},
	T.~Doke\supit{d},		 
        J.~Kikuchi\supit{d}, S.~Ventura\supit{e},  
    	\skiplinehalf
        \supit{a}Columbia Astrophysics Laboratory, Columbia University \\
        \supit{b}Rice University \\
        \supit{c}University of New Hampshire \\
        \supit{d}Waseda University, Japan \\
	\supit{e}INFN and Universit\`a di Padova, Italy
} 

\authorinfo{Further author information: (Send correspondence to: Elena Aprile)
            \\Columbia Astrophysics Laboratory, Columbia University, 
            New York, NY, USA \\ 
            E-mail: age@astro.columbia.edu \hfill}

\begin{document} 
  \maketitle

\begin{abstract}
LXeGRIT is the first prototype of a novel concept of Compton telescope, based on
the complete 3-dimensional reconstruction of the sequence of interactions of
individual \g-rays in one position sensitive detector. This balloon-borne
telescope consists of an unshielded time projection chamber with an active
volume of 400~cm$^2 \times 7$~cm filled with high purity liquid xenon. Four VUV
PMTs detect the fast xenon scintillation light signal, providing the event
trigger. 124 wires and 4 anodes detect the ionization signals, providing the
event spatial coordinates and total energy. In the period 1999 -- 2001, LXeGRIT
has been extensively tested both in the laboratory and at balloon altitude, and
its response in the MeV region has been thoroughly characterized. Here we
summarize some of the results on pre-flight calibration, event reconstruction
techniques, and performance during a 27 hour balloon flight on October 4 --
5. We further present briefly the on-going efforts directed to improve the
performance of this prototype towards the requirements for a base module of a
next-generation Compton telescope.
\end{abstract}


\section{The TPC Approach to a Compton Telescope} \label{sec:TPC_approach}
Between the energy regimes of photoabsorption and pair production, about
250~\keV\ and 6~\MeV\ in xenon, \g-rays interact in matter
predominantly via the Compton process. A Compton telescope is thus the most
promising approach to measure efficiently the distribution of cosmic sources
emitting in the MeV energy band. In a double scatter telescope such as
COMPTEL~\cite{VSchon:93}, the restriction on the event topology to two
interactions recorded in two separate detector planes, results in a relatively
low detection efficiency ($\ll 1\%$).  This figure is greatly improved in a
homogeneous, self-triggered, three-dimensional (3D) position sensitive detector
such as a liquid xenon time projection chamber (LXeTPC), as proposed by
Aprile~et~al.\cite{EAprile:89:ITNS}. In a TPC of large sensitive volume, passive
materials are minimized and a variety of different event topologies are used for
imaging, substantially increasing detection efficiency. With the 3D position
sensitivity, single-site events are easily identified and rejected as
background, as they are not useful for imaging. Multiple-site events are
associated with MeV \g-rays from both source and background. As for any Compton
telescope, the imaging capability greatly enhances signal over background
compared to formerly employed simple photon counters. In addition, the wealth of
information recorded allows for more sophisticated event selections based on
Compton kinematics, interaction locations, or interaction separation.

In absence of a time-of-flight measurement, the correct order of the multiple
interactions has to be determined from the redundant kinematical and geometrical
information measured for each event. The efficiency of this sequence
reconstruction is directly determined by the accuracy of the energy and position
measurements. As previously shown in Aprile~et~al.\cite{EAprile:93:monte_carlo}
and more recently by other groups
\cite{GJSchmid:99:tracking,SBoggs:00:comp_reconst,UOberlack:00:spie00:CSR}~,
event reconstruction based on Compton kinematics allows to correctly order the
interactions and to discriminate against background, e.g., from non-Compton
sequences or multiple coincident gamma-rays.

The LXeGRIT is the outcome of a systematic program initiated at Columbia
University with NASA support to develop the LXeTPC technology for MeV
astrophysics.  Experiments with LXe detectors were carried out to measure charge
and light yields, energy and position resolution, electron drift velocity and
mobility, etc. \cite{EAprile:91:performance}~.  Following these basic studies, a
LXeTPC, with a sensitive volume of 2800 cm$^3$, was developed and its
performance established with a varied of \g-ray sources. The same TPC is used
for the balloon-borne LXeGRIT instrument. 

The detector works as follows. Both ionization and scintillation signals
are detected to measure the event spatial coordinates and the energy. The fast
UV (175~nm) Xe light is detected by four PMTs viewing the sensitive volume from
below, through quartz windows. The OR of the PMT signals is the TPC trigger
signal, marking time zero of each event.  The charge signals induced by free
electrons drifting under the applied electric field are detected on two
orthogonal planes of parallel wires, providing X-Y position information with
millimeter accuracy. The wires pitch is 3~mm, and the separation between X and
Y planes is also 3~mm.  Below the wires, at a distance of 3~mm, four anodes are
used to collect the total charge liberated in the event. The charge is directly
proportional to energy.  The 62 X-wires and 62 Y-wires and the anode are
amplified and digitized at 5~MHz, with 8 bit precision for the wires and 10 bit
precision for the anodes.  The digitized data are stored with 256 samples per
event, covering more than the maximum drift time of 35 $\mu$s.  The absolute
drift time measurement gives the event depth of interaction (Z-position), with
an accuracy of 300 $\mu$m.  This charge readout was designed and tested to image
the point-like ionization clouds ($<$~1mm) produced by low energy Compton
electrons and photoelectrons which are typical of MeV \g-ray interactions in the
dense LXe. The signal size associated with these clouds is very small (W=15.6~eV
compared to W=2.96~eV for Ge and W=3.62~eV in Si). A key prerequisite for this
detector to work is to minimize the signal loss due to electron trapping by
impurities in the liquid.
In LXeGRIT free electrons are drifted over the TPC maximum distance of
7~cm with very little attenuation, and the remaining 5\% maximum charge loss for
full drift length is corrected for based on the known Z-position. 

The TPC is enclosed in a cylindrical vessel filled with 7 liters of pure LXe
kept at $\sim -95\deg$C by a controlled flow of LN$_2$ through a condenser. The
vessel is thermally insulated with a vacuum cryostat. The lower section of the
cryostat houses the four PMTs of the xenon light readout, as well as the HV
distribution circuitry for the wires and the cathode.

\section{The LXeGRIT Balloon Flight Program} \label{sec:flight}

To demonstrate the operation and performance of the LXeTPC with \g-rays in the
near space environment, the detector was turned into a balloon
payload called LXeGRIT. 
This instrument is the first prototype of a Compton telescope using a single, 3D
position sensitive detector like a LXeTPC. 
Having shown its \g-ray imaging capability in the controlled environment of the
laboratory, it was clear that the practical implementation of this
new type of instrument for astrophysics would present a number of challenges
which needed to be directly tested with a balloon flight.   
For this purpose, all detector subsystems had to be capable
of sustaining the near vacuum conditions and the temperature extremes
encountered during a balloon flight. 
In addition, two major new developments were required to make a balloon
experiment with the LXeTPC: a low-power readout electronics and data acquisition
flight system, and an instrumentation and control system. These systems were
developed in close collaboration and with engineering support from Marshall
Space Flight Center and are described in
Aprile~et~al.\cite{EAprile:98:electronics}~.  The front-end electronics, also
described in \cite{EAprile:98:electronics}, was developed in collaboration with
the Waseda group in Japan and the Clear Pulse Company. The hardware
modifications to the original systems, and the new data acquisition software
which followed the 1997 flights, are discussed in
Aprile~et~al.\cite{EAprile:00:spie00:performance}~.  A parallel effort involved
extensive modifications of an existing gondola and veto shield systems provided
by the University of New Hampshire. Being a Compton telescope, LXeGRIT does not
require an active pointing system. The instrument's zenith direction and azimuth
orientation is provided by a sensor combining 3-axis magnetometer and 3-axis
accelerometer. This information, together with the knowledge of the payload
geographical coordinates, is needed for imaging analysis of celestial sources.
The flight data are stored in two 36-G-byte disk drives and also sent via
telemetry at $2 \times 500$~kbps to the ground station. 
Table~\ref{t:lxegrit} summarizes the instrument characteristics.

\begin{table}[htb]
\vspace{\baselineskip}
\centering \sloppy 
\abovecaptionskip 5pt
\tabcolsep 2pt
\begin{tabular}{|p{0.02\linewidth} p{0.35\linewidth} |p{0.02\linewidth}
p{0.32\linewidth}|}\hline 
\tabcolsep 2pt
\raggedright
& Energy Range & & $0.15$ -- $10$ MeV\\
& Energy Resolution $(FWHM)$  & & $8.8\%\times(1\,\mathrm{MeV}/E)^{1/2}$ \\
& Position Resolution $(1\sigma)$ & & 1 mm (3 dimensions)\\
& Angular Resolution $(1\sigma)$ & & 3.8\deg\ at 1.8~\MeV \\
& Field of View (FWHM) & & 1~sr \\
& Detector Active Volume & & 20 cm $\times$ 20 cm $\times$ 7 cm \\
& Instrument Mass,\enspace Power & & 2000 lbs,\enspace 450 W \\
& Telemetry,\enspace On-board Data Storage & & $2 \times 500$~kbps,\enspace 
	                                    $2 \times 36$~GB \\
\hline
\end{tabular}
\caption{LXeGRIT characteristics in the balloon flight 2000 configuration.}
\label{t:lxegrit}
\end{table}

\subsection{Pre-flight Calibration} \label{sec:calibration}

\begin{figure}[htb]
\centering
\medskip
\medskip
\includegraphics[bbllx=77,bblly=60,bburx=580,bbury=380,
width=0.55\linewidth,clip]{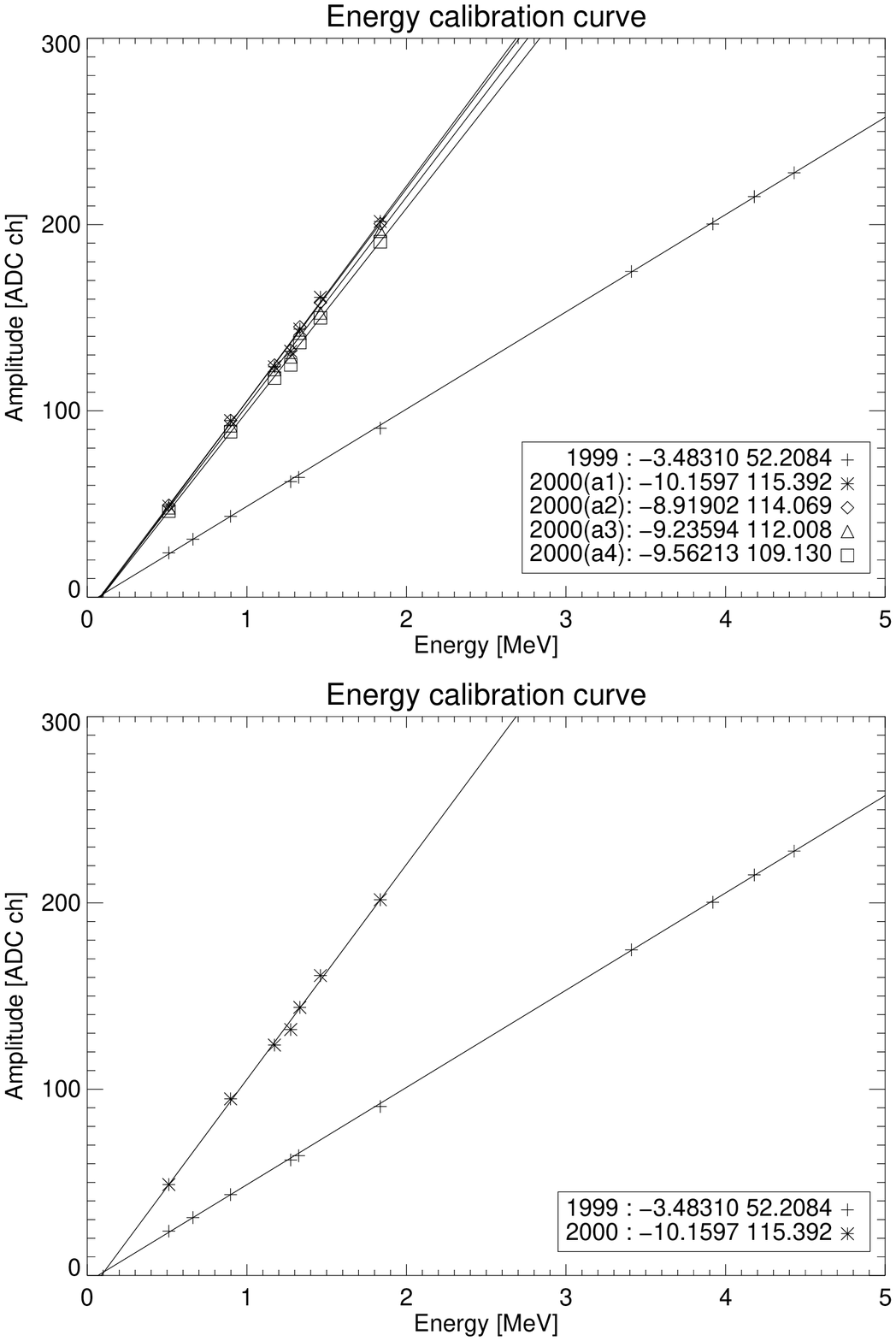}
\caption{LXeGRIT energy calibration curve for one of the four anodes from '99
({\it crosses}) and 2000 ({\it stars}) pre-flight calibration data.} 
\label{f:fig_cal}
\end{figure}

\begin{figure}[h]
\centering
\includegraphics[bbllx=50,bblly=60,bburx=300,bbury=267,%
	width=0.5\linewidth,clip]{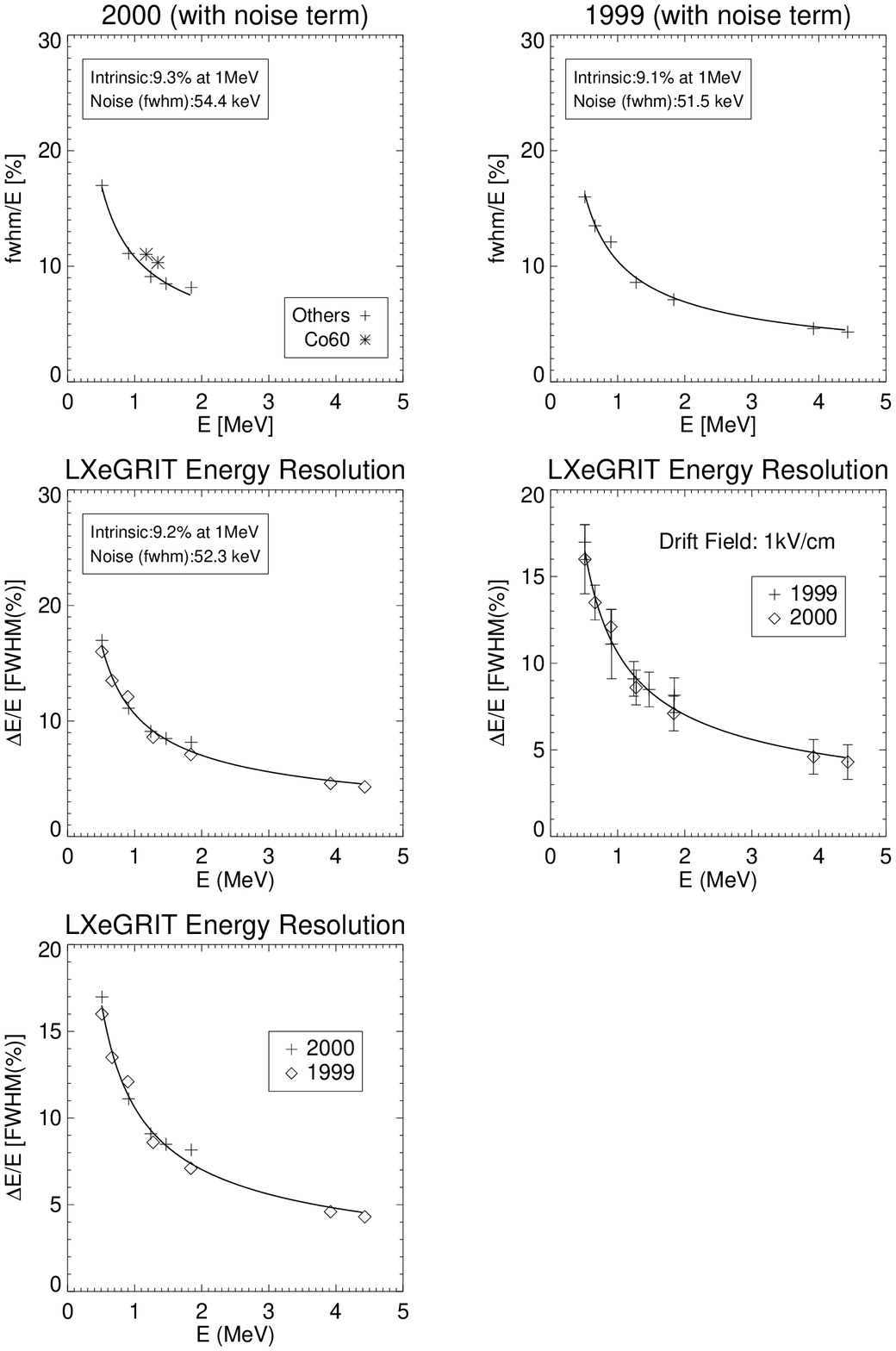}
\caption{LXeGRIT energy resolution {\it vs.} energy, at 1~kV/cm, as measured
from '99 ({\it open diamonds}) and 2000 ({\it crosses}) pre-flight calibration
data.}  
\label{f:elec}
\end{figure}

Results from the 1999 pre-flight calibration have been reported in
Aprile~et~al.\cite{EAprile:00:spie00:performance}~. 
The 2000 pre-flight calibration gives a similar spectral and spatial resolution
performance, but differs in trigger efficiency and data acquisition (DAQ) speed. 
The electronic gain on the anode signal was increased by a factor of $\sim$2
(see Fig.~\ref{f:fig_cal}), reducing the maximum energy before saturation of the
FADC dynamical range from $\sim$20~\MeV\ in 1999 to $\sim$10~\MeV\ in
2000, which is a more appropriate upper limit for this experiment. \\
The intrinsic energy resolution at a drift field of 1~kV/cm
is consistent with $ \Delta
E_\mathrm{lxe}/E=8.8\% \: \sqrt{1\MeV /E}$ (FWHM) and it is linear for \g-rays
in the energy range 0.5-4.4~\MeV\ (see Fig.~\ref{f:elec}), in accordance with
results from small ionization chambers.  
The contribution from electronic noise is about 55~\keV\ (FWHM).


\begin{figure}[htb]
\centering
\medskip
\medskip
\hspace{1.0cm}
\includegraphics[bbllx=55,bblly=508,bburx=290,bbury=711,%
	width=0.40\linewidth,clip]{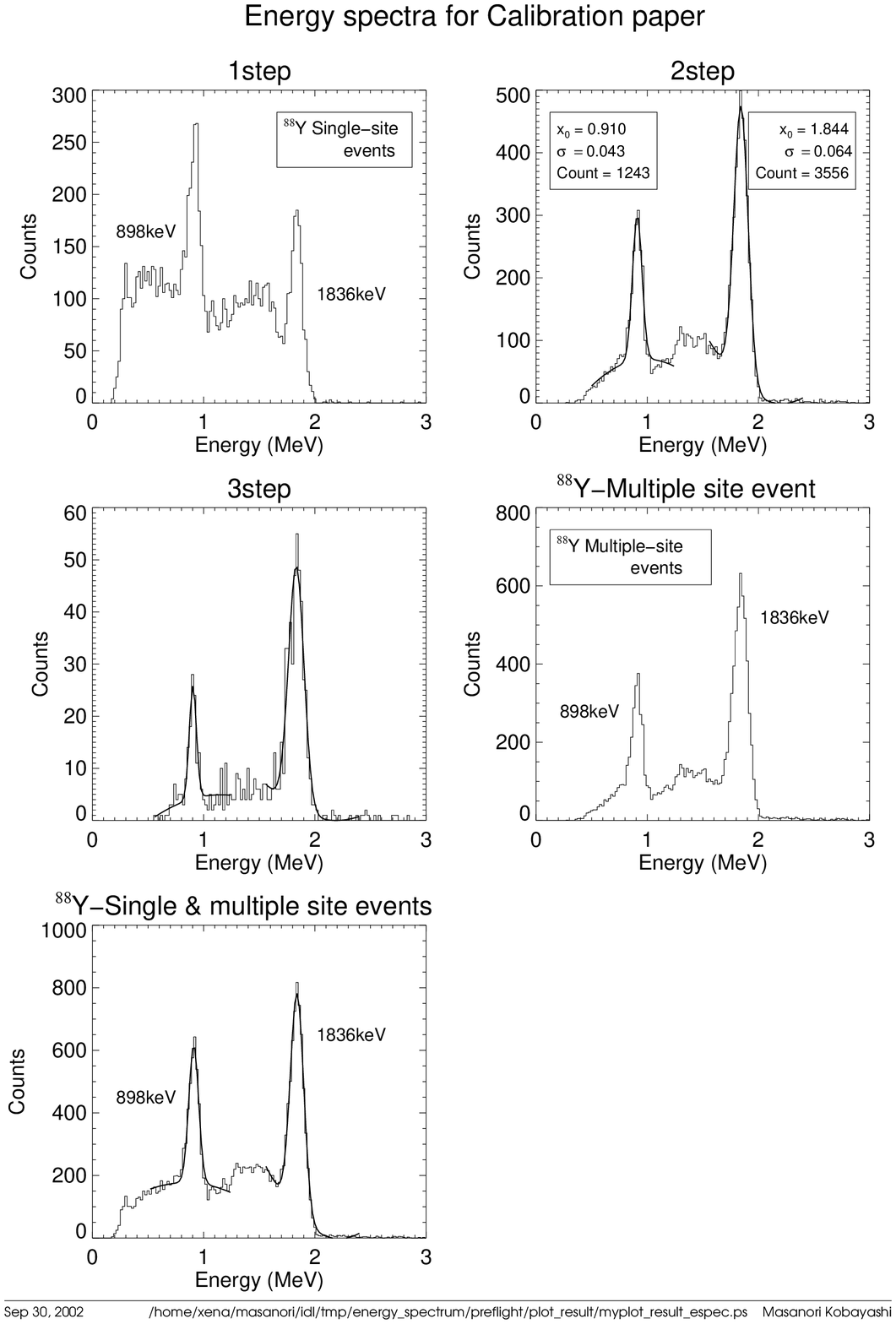}
\hfill
\includegraphics[bbllx=300,bblly=280,bburx=535,bbury=483,%
	width=0.40\linewidth,clip]{ypa4_espec_new.ps}
\hspace{1.0cm}

\bigskip
\bigskip
\hspace{1.0cm}
\includegraphics[bbllx=55,bblly=508,bburx=290,bbury=711,%
	width=0.40\linewidth,clip]{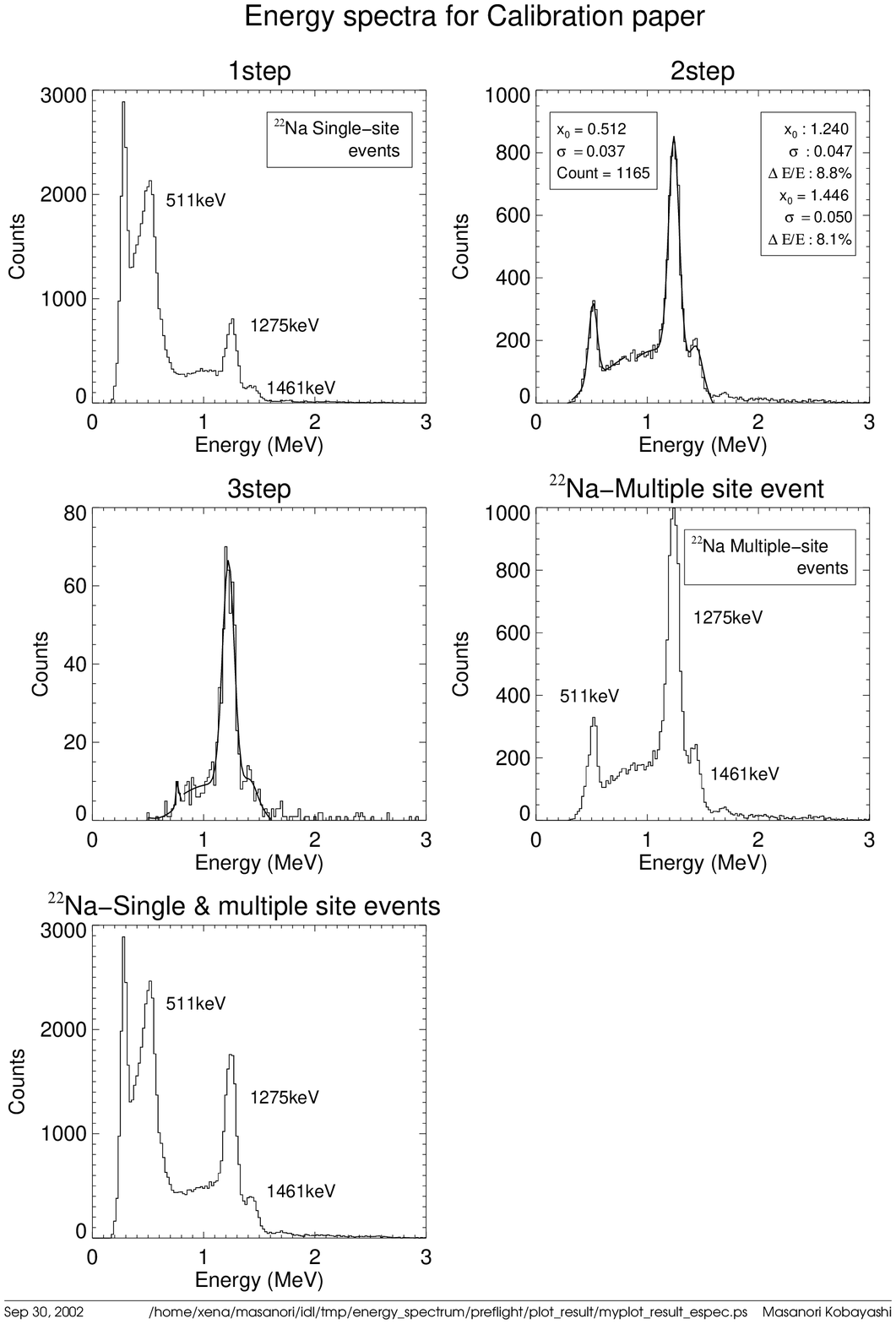}
\hfill
\includegraphics[bbllx=299,bblly=282,bburx=535,bbury=485,%
	width=0.40\linewidth,clip]{napa_espec_new.ps}
\hspace{1.0cm}
\caption{{\it Top left:} energy spectrum for single-site events from an \yt\
source (898 and 1836~\keV\ lines). {\it Top right:} multiple-site events, \yt,
same experiment. 
{\it Bottom left:} energy spectrum for single-site events from a \na\ source
(511 and 1275~\keV\ lines). {\it Bottom right:} multiple-site events, \na, 
same experiment. The line at 1465~\keV\ ($^{40}$K) is due to natural
radioactivity of the ceramic (MACOR, which contains about 10\% potassium) used
around the wire structure and field shaping rings.} 
\label{f:ypa4_napa}
\end{figure}

The large amount of information made available for each \g-ray event by the
fine granularity of the TPC allows a detailed spectral analysis. The precise
knowledge of the event topology suggests a separate analysis for events with
different interaction multiplicity. 
Also due to the relatively large size of the LXeTPC, for \g-ray energies larger
than $\sim$800~\keV, the detection efficiency is significantly higher for
multiple-site events than for single-site events.
Summing up the separate interactions of multiple-site events, 
a largely enhanced {\it peak-to-Compton} ratio is also obtained, as shown in
Fig.~\ref{f:ypa4_napa} for an \yt\ (898 and 1836~\keV\ lines) and a \na\ source
(511 and 1275~\keV).

%

The Compton Sequence Reconstruction (CSR) is based on the testing of Compton
kinematics, using the redundant information available in the measurement of
positions and energy depositions in three or more interactions. We currently
employ a test statistic which measures the differences square of the cosines of
the geometrically measured angles \phigeo\ and the photon scattering angles
\phibar\ inferred from the measured energy depositions. This sum is weighted by
the measurement uncertainties in both quantities, $\sigma_{\cos{\phibar}}$ and
$\sigma_{\cos{\phigeo}}$:    
\begin{equation} \label{e:test:CSR}
  T = \frac{1}{N-2} \: 
      \sum_{i=2}^{N-1} \frac{(\cos{\phibar_i} - \cos{\phigeo\ _i})^2}%
                            {\sigma_{\cos{\phibar},i}^2 +
                             \sigma_{\cos{\phigeo},i}^2  }
\end{equation}
where $N$ is the number of interactions in the detector.  Ideally, the test
statistic would be zero for the correct sequence if the photon is fully
contained. With measurement errors and a contribution from Doppler broadening,
resulting from the binding energy of the interacting electron, $T$ is larger
than zero. Even so, the correct interaction sequence most likely produces the
minimum value of the test statistic among all possible sequences. In addition to
minimizing $T$, an upper threshold on $T$ can help in discriminating against
photons that are not fully absorbed or that interact through mechanisms other
than Compton scattering. Details can be found in
Oberlack~et~al.\cite{UOberlack:00:spie00:CSR}~.

\begin{figure}[htb]
\belowcaptionskip 0pt
\centering
\includegraphics[bb=61 276 293 490,height=0.28\textheight,clip]{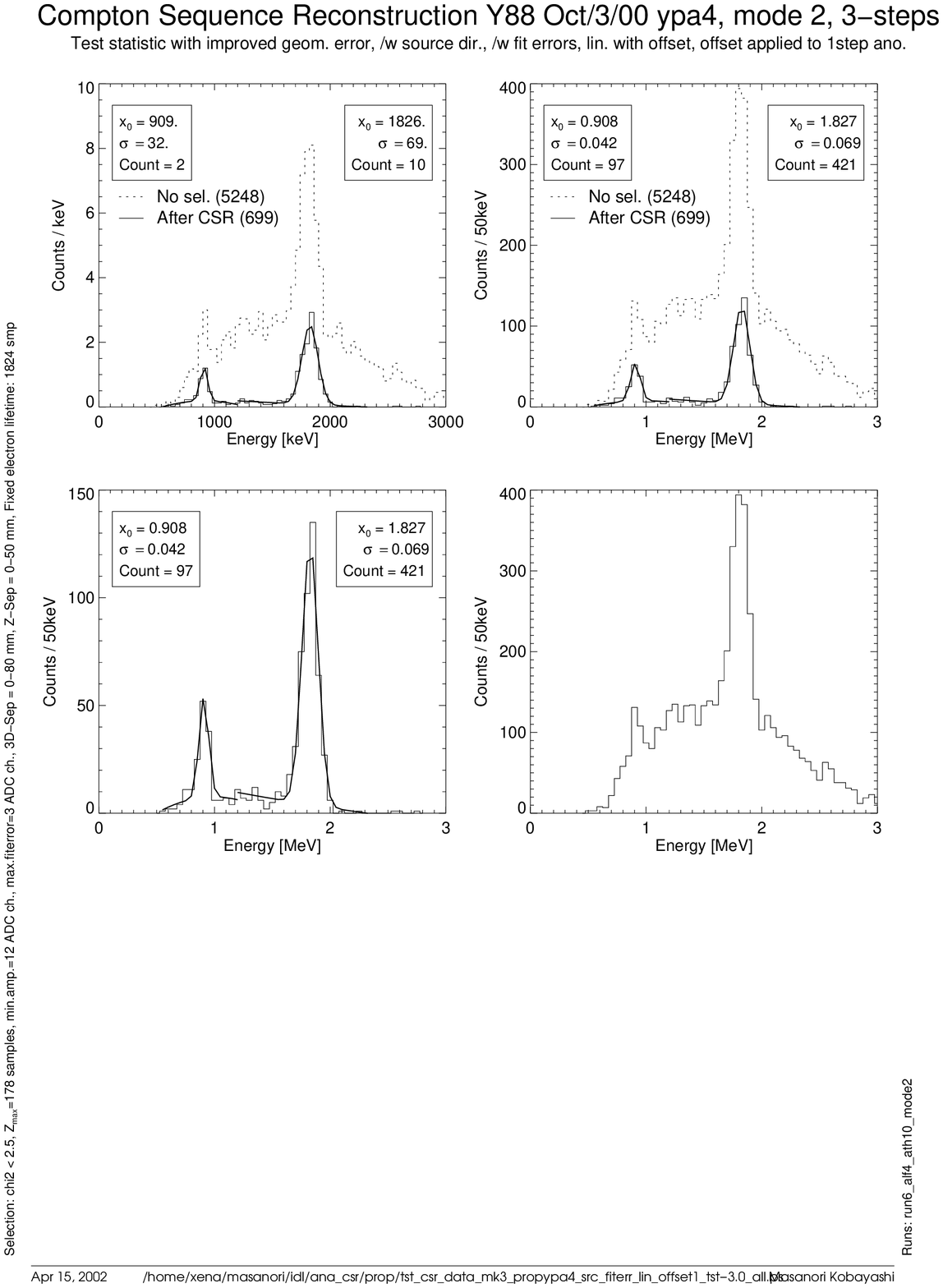}
\caption{\yt\ ``3--site events'' energy spectrum, after event selections and
CSR} 
\label{f:y_spect}
\medskip
\includegraphics[bb=55 500 297 727, height=0.28\textheight,clip]{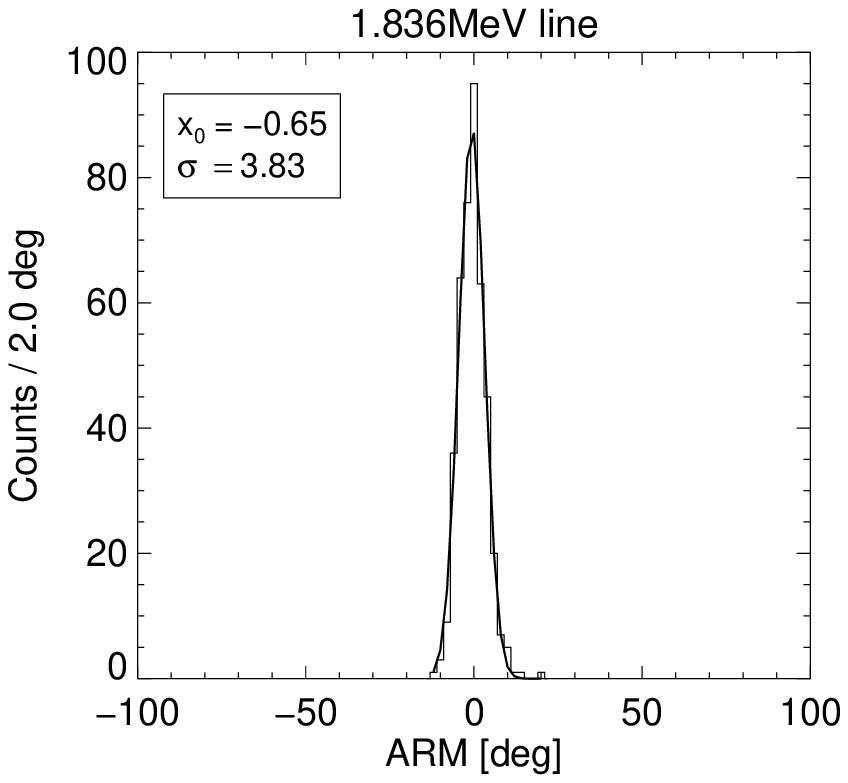}
\caption{Measured angular resolution distribution for 1.8~\MeV}
\label{f:ARM:y88}
\end{figure}

The CSR algorithm has been applied to multiple-site events from \na, \co\ and
\yt\ sources placed a few meters above the LXeTPC.  Event selections made
possible by 3D position resolution and CSR result in a dramatic suppression of
background, as shown in the clean energy spectrum of \yt\ obtained with 3-site
Compton events (Fig.~\ref{f:y_spect}).
For photons losing energy in two interactions only, the problem as in
Eq.~\ref{e:test:CSR} is underconstrained, but probabilistic approaches can still
be used, which turned out quite effective especially for \g-ray energies above
1~\MeV. 

Once the correct interaction sequence has been found we can image \g-ray
sources. Imaging proceeds as for a classical double scatter Compton telescope,
i.e. by determination of the scatter direction between first and second
interaction and by the measurement of the Compton scatter angle $\varphi$, based
on measured energy deposits and Compton kinematics.  The measured scatter angle
define a circle on the sky, on which the \g-ray source is located. Many photons
from the same source result in intersecting ``event circles'' with varying
radii. The intersection point defines the source position. Additional
information is available from the known probability density for scatter angles
within the telescope for a given source location. This leads to more powerful
imaging techniques based on likelihood analysis. Given the necessary
event-by-event analysis, it is efficient to employ so called list-mode imaging
techniques \cite{HBarrett:97:listmode}, where the probability for a photon to
originate in a certain point in the sky is determined for each event, rather
than in a binned dataspace as used by COMPTEL. We have implemented imaging
software and tested the method successfully on calibration data. An in-depth
discussion of these results will be reported elsewhere.

\begin{figure}[htb]
\centering
\includegraphics[bb=263 76 492 362,width=.4\linewidth,angle=90,clip]{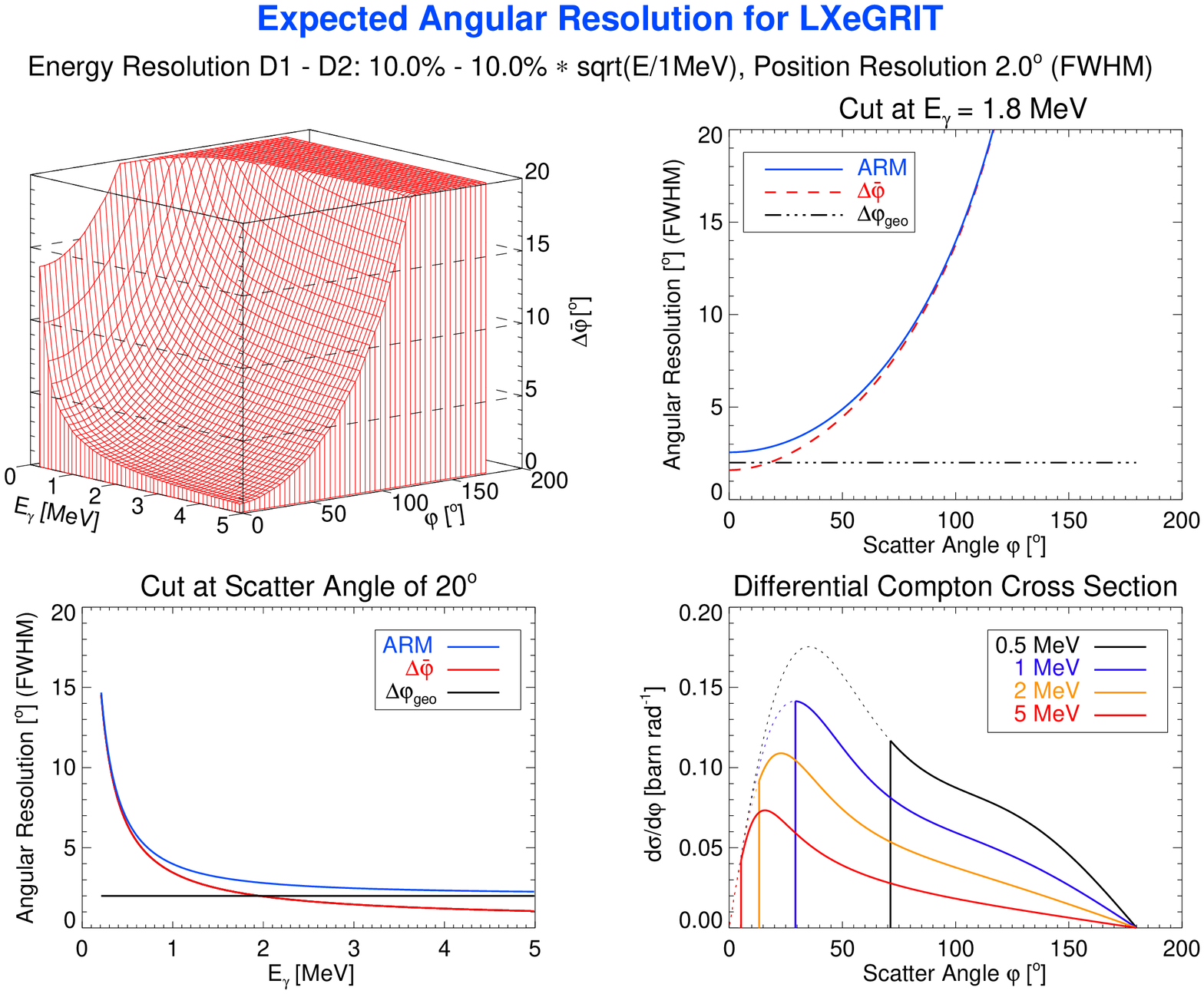}  
\caption{\label{f:angres_vs_phi} Expected angular resolution {\it vs.} \phibar
\ for 1.8~\MeV \ photons (solid line) and contributions from energy resolution
(dashed line) and position resolution (dash-dotted line).} 
\end{figure}

The angular resolution of a Compton telescope is given by the ``angular
resolution measure'' $ARM = \phibar - \phigeo$.  Fig.~\ref{f:ARM:y88} shows the
measured $ARM$ distribution at 1.8 MeV, consistent with the expectation (see
Fig.~\ref{f:angres_vs_phi} for comparison).  Clearly, events with smaller
scatter angle would result in better angular resolution.
   
The detection efficiency for Compton events has been both directly measured and
studied through Monte Carlo simulations taking into account the LXeGRIT mass
model, minimum energy threshold and energy and spatial resolution. It varies
between 1.5~\% and 4~\% depending on energy and event selection, which
corresponds to a quite large effective area for this kind of detector, from 6 to
16~cm$^2$. This efficiency does not account for the limited DAQ speed (see
Aprile~et~al.\cite{EAprile:01:daq}) and inefficiencies of the light-trigger,
which in the end dominate the total efficiency.   

As discussed in Oberlack~et~al.\cite{UOberlack:01:trigger}~, for the 2000 flight
we chose to {\it reduce} the trigger efficiency to specifically select multiple
Compton events in the few \MeV\ region and to be less sensitive to lower-energy
gamma-rays, given the dead-time limited DAQ.
The light-trigger efficiency was measured for energies up to 2~\MeV\, spatially
resolved with fine granularity. From these data, a detailed model of the
detector trigger efficiency was derived.    

Another handle we have to select multiple-site events is to reject single-site
events on-line. This selection, based on the number of wire hits, is implemented
in the LXeGRIT data acquisition software. This selection is very powerful, with
very low acceptance of single-site events at energies below 2~\MeV\ and about
50~\% above 5~\MeV, where the fraction of single-site events is anyway small
\footnote{
More details about the LXeGRIT efficiency in flight 2000 configuration are given
in Curioni~et~al. ``On the Background Rate in the LXeGRIT Instrument during the
2000 Balloon Flight''\cite{ACurioni:02:bkgd}}~.


\subsection{In-flight Performance} \label{sec:inflight}

\begin{figure}[ht]
\centering
\includegraphics[bb=69 36 552
731,width=0.69\textwidth,clip]{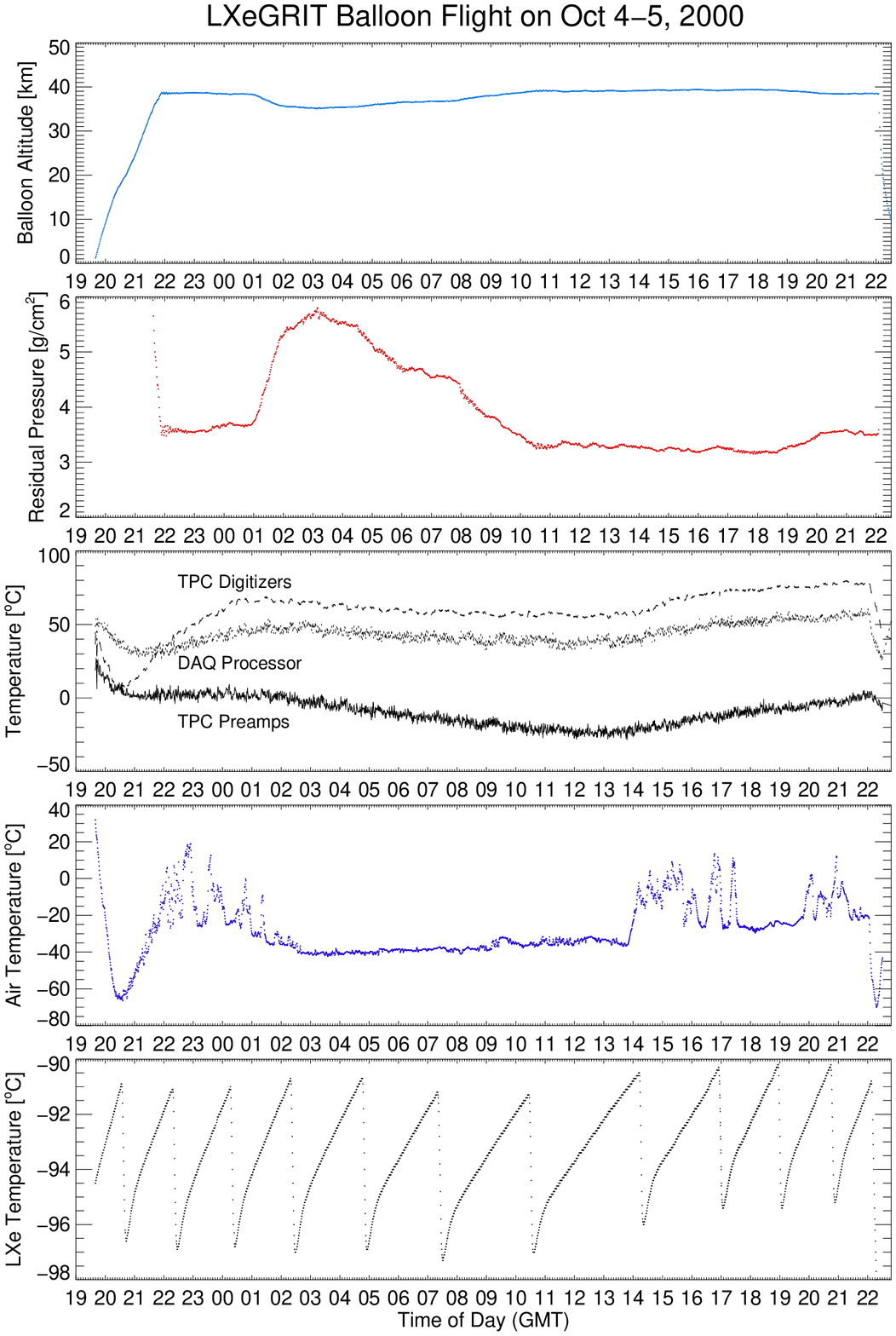} 
\caption{{\it From top to bottom:} 
1. balloon altitude during flight;
2. corresponding atmospheric depth;
3. temperatures as measured by sensors placed at three different locations
   (TPC preamplifier boxes, DAQ processor and TPC electronics board); 
4. air temperature - some apparently wild variations are due to direct
   exposure to the Sun as the gondola rotates around its vertical axis;
5. liquid xenon temperature - the ``dips'' correspond to cooling cycles with
   liquid nitrogen.}
\label{f:flight_HK}
\end{figure}

A primary goal of the LXeGRIT balloon experiment was to measure the background
rate at float altitude for this novel instrument. 
%
Since the maiden engineering flights from Palestine, TX, in the Summer of 1997,
LXeGRIT has been improved and has successfully flown again twice from Ft.Sumner,
NM, in May 1999 and in October 2000.  
The improvements were focused on the DAQ and trigger system, onboard data
storage and telemetry, as well as on instrument support systems
\cite{EAprile:00:spie00:performance,EAprile:01:daq,EAprile:01:pisa}~. 
The TPC itself was never modified and has required only minimal repairs during
the five years since 1997.  The 1999 flight lasted almost 10 hours, while the
2000 flight lasted 27 hours, including two hours ascent, with 40~GB of data
collected.  It was launched on October 4 from the National Scientific Balloon
Facility in Fort Sumner, New Mexico, at 19:39:48~UT and the detector operation
was stable throughout the flight. The electronics, cryogenics system and DAQ
performed as expected from testing in the laboratory. Some relevant parameters
describing the flight conditions are summarized in Fig.~\ref{f:flight_HK}. The
payload was recovered in good conditions 10 miles south of Buckeye, Arizona. 

A detailed description of LXeGRIT in the 1999 flight configuration is given in
Aprile~et~al.\cite{EAprile.2000b.SPIE}~, which can be considered as a reference
also for the 2000 flight configuration but for three relevant points:
\begin{enumerate}
\item the active NaI(Tl) and liquid scintillator shields, surrounding the
      LXeTPC, were removed to better understand the response of the 
      TPC itself to space radiation. The plastic scintillator charge particle
      shield above the TPC was also removed;
\item a gain of a factor 2.5 in DAQ speed (see
      Aprile~et~al.\cite{EAprile:01:daq}) and a factor of about 2 in inflight
      event data transfer to ground;
\item optimization of the light trigger and on-line selections for \MeV\
      multiple-site events (see Oberlack~et~al.\cite{UOberlack:01:trigger} and
      Curioni~et~al., accompanying paper in this proceeding). 
\end{enumerate}

\begin{figure}[ht]
\centering
\psfig{file=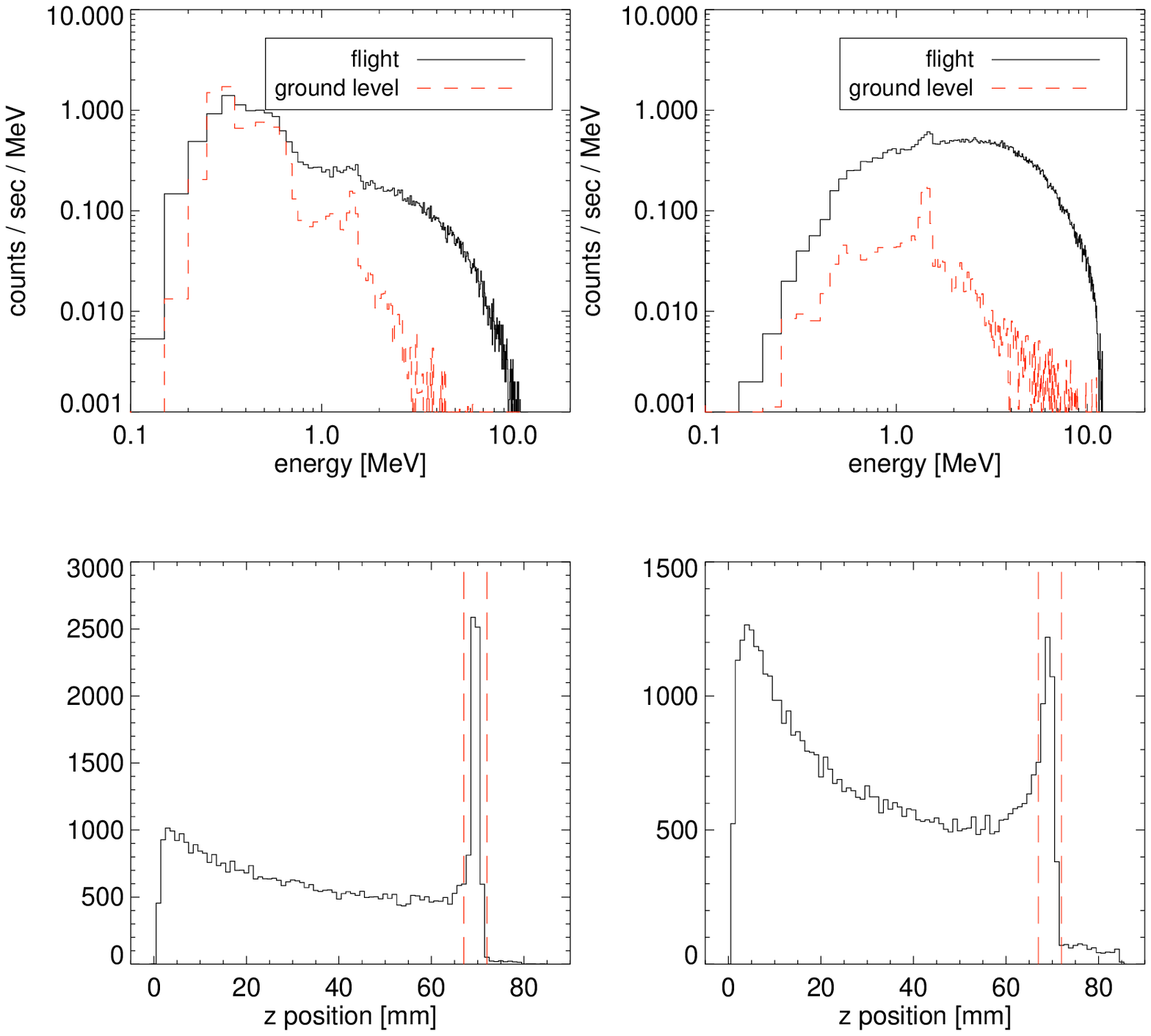,bbllx=84,bblly=515,bburx=544,bbury=718,width=.85\linewidth,clip=} 
\caption{\label{f:intbkgd_2} Energy spectra for flight data (\emph{continuous
line}) and for a background run at the ground level (\emph{dashed line}). {\it
Left:} single site events. \emph{Right:} multiple interaction events.}
\end{figure}

The energy spectrum of single- and multiple-site events recorded at balloon
altitude in the period 12:00 - 17:00~UT is shown in Fig.~\ref{f:intbkgd_2}. As
discussed in Curioni~et~al. in this proceeding, the measured spectrum is
reproduced within a factor of 2 with atmospheric and cosmic diffuse \g-ray
fluxes, a mass model of the payload, and a model of the internal background as
measured on the ground, suggesting that activation of passive materials due to
atmospheric neutrons and protons is only a small contributor in this unshielded
detector.  The instrument has been modeled using the GEANT package and all the
required instrument parameters (energy thresholds and resolution, spatial
resolution, light-trigger efficiency etc.) have been measured independently,
while the livetime fraction of the DAQ system is computed on-line.

The 1.46 MeV line, clearly visible in the multiple-site events of Fig.~\ref{f:intbkgd_2}, is attributed to natural $^{40}$K
radioactivity, mostly from the machinable ceramic (MACOR), used to support the
TPC wire structure and the field shaping rings. Macor contains about 10~\%
potassium oxide.  The $^{40}$K line and most of the continuum below 500~\keV, is
present at a similar rate in the LXeGRIT background spectrum measured on the
ground.
Activation of the xenon itself also appears negligible, as we do not detect any
other line. This observation is consistent with laboratory experiments with LXe
detectors exposed to neutron beams and experiments in deep space (see
Kirsanov~et~al.\cite{MAKirsanov:93} and Ulin~et~al.\cite{SEUlin:98}). 

As reported previously\cite{EAprile.2000b.SPIE}~, the number of multiple-site
events recorded during the 1999 balloon flight was largely reduced compared to
single-site events. The light trigger threshold was set very low; the thick NaI
shields below and around the TPC were supposed to veto a large fraction of this
component. Once afloat, the trigger rate turned out unexpectedly large and the
trigger upper discriminator threshold was lowered to reduce the DAQ dead-time,
cutting into the \MeV\ band.  
Analysis of the data and Monte Carlo simulations with atmospheric neutron flux
indicate that a combination of neutron induced background from shields
activation, only partially working shield sections, and the enhanced trigger
efficiency at low energy, may explain the measured background rate and spectral
shape.

\begin{figure}[htb]
\bigskip
\bigskip
\centering
\hspace{1.0cm}
\includegraphics[bb=40 510 298 717,%
	width=0.4\linewidth,clip]{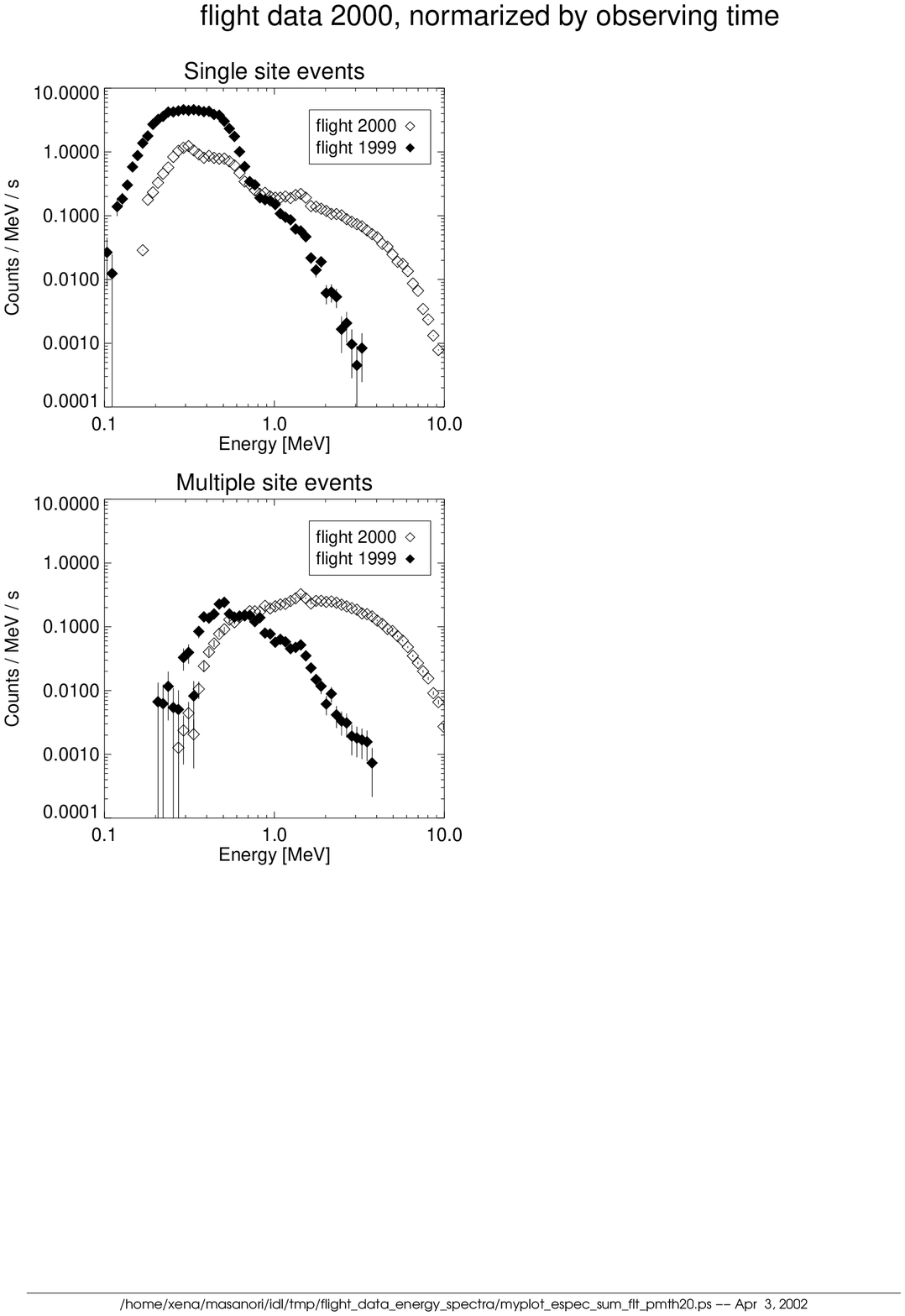} 
\hfill
\includegraphics[bb=40 283 298 490,%
	width=0.4\linewidth,clip]{espec_flight99and00vlog.ps} 
\hspace{1.0cm}
\caption{In-flight energy spectra, normalized to the total exposure but not
corrected for DAQ livetime:~1999 and 2000. {\it Left:} single-site events, {\it
right:} multiple-site events. }  
\label{f:espec:flight}
\end{figure}

For the 2000 flight we removed all the shields and optimized the detection
efficiency for multiple-Compton events in the few \MeV\ energy region. 
The trigger rate as given by the PMT-OR was relatively low, $\sim$600~Hz
and nearly constant throughout the flight, after ascent had been
completed. Albeit the PMT-OR gave a much lower rate compared to the one during
the 1999 flight, the final rate of \MeV\ \g-ray events after selections is much
higher for the 2000 flight settings.
A comparison between the two flights is shown in Fig.~\ref{f:espec:flight} where
the rate is normalized to the total exposure, but not corrected for DAQ
livetime. The much improved efficiency in the \MeV\ region and the increased
number of multiple-site events in the 2000 flight data is self-evident.  

Beyond establishing the in-flight background spectrum in an LXe Compton
telescope, we would like to prove its imaging performance and background
suppression capability by identifying the signal from the Crab nebula. We have
in fact seen hints for a signal of the proper strength in two independent data
samples, based on 2-site and 3-site events. ARM spectra show an excess at ARM =
0\deg, which is not present in background spectra.  We are still studying the
impact of event selections on signal/background during the flight. Our current
efforts are pointed towards identifying the data cuts that best reject
background without further reducing the limited number of source counts. This
includes careful studies of the events we have rejected in a first analysis
based on event quality, a study of fiducial volume and spatial separation cuts,
optimized also with the help of Monte Carlo simulations, and improving the
algorithms to identify the proper interaction sequence. We are also working to
improve background models for a likelihood imaging analysis, and we plan to
search for timing structure in our data from the Crab pulsar. The results of
this ongoing effort are beyond the scope of this paper and will be reported
elsewhere. 

\section{Future Prospects: LXeGRIT-2 and Beyond} \label{sec:future}

Following the 2000 flight, several experiments with the TPC have been carried
out for post-flight calibration and for studies of the light trigger
system. From the extensive set of measurements, data analysis results, and
simulations, we have identified strengths and weaknesses of the
current prototype design. While various goals of basic research and development
remain to explore the full potential of this technology, we consider
the LXeTPC technology mature enough to allow implementation of a large
instrument in the near future. We will summarize in the following two sets of
improvements that would vastly boost the efficiency and overall performance of
the current instrument. The first set, described in Sec.~\ref{sec:lxeg2}, could
be implemented as a very significant upgrade of the current system, which we
call LXeGRIT-2.  The second set would go even further, towards the development
of a base module for a next-generation Compton telescope based on LXeTPC
technology.

\subsection{LXeGRIT-2} \label{sec:lxeg2}

The following measures have been identified for upgrading  the
current detector, largely maintaining the current read-out system. We have
started to implement a new light trigger system and the other changes have been
studied.
\begin{enumerate}
\item  Install a new light detection system based on 12 Hamamatsu R6041
phototubes in the liquid surrounding the sensitive volume from all sides. 
The sum signal of the 12 PMTs, followed by the window discriminators already
available, will provide the TPC trigger. 
\item Replace the cathode and HV connections to operate the TPC at 3~kV/cm
electric field. Move the front-end amplifiers inside the cryostat in close
proximity to the signal feedthroughs. Minimize analog pre-filtering before
digitization on the anodes.
\item Add two more readout processors in order to read the X- wires, the Y-
wires, and the anodes in parallel.
\item Add a plastic scintillator shield to veto charged particles at the trigger
level. 
\item Add a Global Positioning System (GPS) interface
for precise event time stamping, needed for pulsar studies.
\end{enumerate}

The first modification is to enhance the energy sensitivity of the trigger and
thus reject background outside the energy band of interest without requiring
action of the readout processor. The current geometry, with four external PMTs,
gives a poor light collection efficiency and a large, position dependent,
amplitude spread. This requires a very low trigger threshold. Our raw trigger
rate is then much higher due to low energy background and noise pulses. The new
geometry, with 12 PMTs surrounding the TPC sensitive volume, and the addition of
UV reflectors, would give about tenfold increase in light collection, as shown
by ray-tracing simulations. The ability to detect the \g-ray source energy with
the light will allow to select the energy band of interest at the trigger
level. At 1~\MeV, an increase in trigger efficiency from $\sim 10\%$ to $\sim
100\%$ is expected. A direct consequence of the better light collection
efficiency is to improve the energy resolution by adding the scintillation
signal to the ionization signal. This is currently being investigated with a
small ionization chamber equipped with one Hamamatsu R6041.  

The Hamamatsu R6041 PMT is a 2" diameter, metal channel tube. It was a special
development for LXe detectors. We have tested its performance in LXe and studied
its compatibility with the high purity requirement for drifting
charges. Machining of the Teflon (90\% reflectivity in the UV) walls supporting
the PMTs is complete and a new design of anode and 
cathode to increase reflectivity is underway. The new geometry will also
eliminate triggers from the layers of LXe outside the fiducial volume. In the
present chamber, with the PMT's viewing the chamber from below, a 3 cm LXe layer
below the anodes is an efficient scintillator slab, in which background events
easily produce false triggers.

The second modification will bring both better energy resolution and lower
electronic noise, and thus lower energy thresholds. The drift field on the TPC
is currently limited to 1~kV/cm. By replacing the cathode and the HV
connections, we will be able to increase the applied voltage from 9~kV to 25~kV,
needed for a 3~kV/cm drift field. For this increase in applied electric field,
we have previously measured the FWHM energy resolution to improve from 12.5~$\%$
to 7.5~$\%$ for 662~\keV\ (\cs). 

With a clever arrangement of the 128 signal feedthroughs on a new TPC bottom
flange, we will be able to connect the charge sensitive amplifiers with minimum
stray capacitance, and thus reduce the noise. For the anodes the capacitance
will be reduced from the current 76~pF to 24~pF, corresponding to a reduction in
the noise level from 1200 electrons RMS to about 480 electrons RMS. 

With lower noise, analog pre-filtering before digitization, and therefore signal
shaping, can be significantly reduced, minimizing the separability of two
signals on the same anode from the current value of $\sim 4$~mm. This would
further improve the number of optimally reconstructable events, for which wire
and anode signals are fully matched.

The third action of increasing the number of readout processors will bring an
additional increase in overall efficiency. As previously mentioned, the current
DAQ system is about 50$\%$ dead-time limited for a trigger rate of few hundreds
Hz, because of the slow data transfer from the digitizers. Since the on-line
software treats the X-wire, the Y-wire and the anode signals nearly
independently, the time required to process an event is reduced by a factor 2.5
by splitting the tasks and having three  processors work in parallel. All three
processors will run the same software as in the present readout, but each is
connected to only one section  of the digitizer system. Once accepted, the data
will be combined to form a single event and sent either to the telemetry
transmitters or the two hard disks for onboard storage.  

\begin{figure}[htb]
\bigskip
\bigskip
\centering
\includegraphics[width=0.4\textwidth,clip,angle=90]{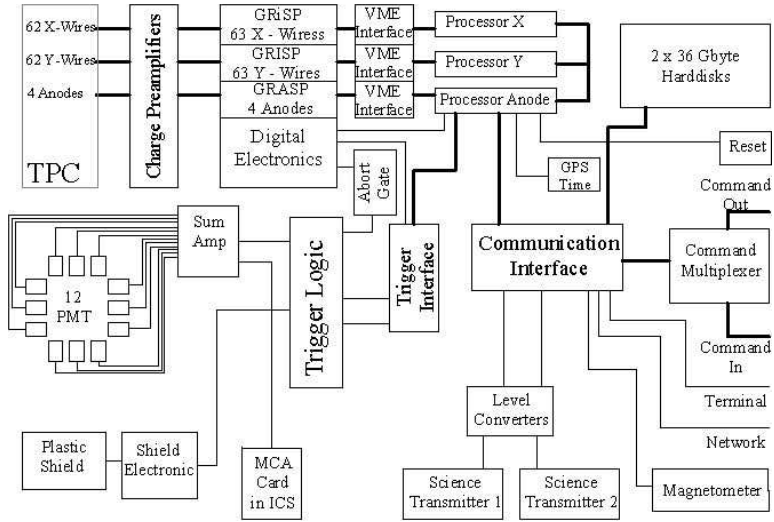}
\caption{Block diagram of the LXeGRIT-2 flight system.}
\label{f:blockdiag}
\end{figure}

Finally, with a plastic shield fully covering the TPC, charged particles will be
vetoed at the trigger level. As we have directly verified, the majority of
charged particles can be easily identified from the large energy detected on the
anodes and rejected by the on-line DAQ software. Particles cutting the corners
deposit much less energy and require more off-line processing to ensure they are
not confused with \g-ray interactions at the edge of the detector. In the
current system, a charged-particle rejection
on the trigger level could reduce the event rate to be handled by the DAQ system
by as much as 2/3.
In 1999, a thin plastic counter was used to cover the TPC aperture, since the
sides and bottom were covered by the UNH NaI shields. In 2000, all shields were
removed. The charged particle rate measured from the in-flight data digitized in
full mode, was about 400 Hz, more than 60$\%$ of the total trigger rate of
$\sim$600Hz. The proposed shield for the next LXeGRIT flight will keep
processing time low by vetoing all charged particles in the trigger.   
Together with the energy sensitivity of the light trigger, we expect to entirely
defeat the present dead-time limited behavior of the DAQ system, and effectively
increase the amount of science data by a factor two for the same flight time
(Fig.~\ref{f:blockdiag}). 

A commercial GPS together with a time interpolator would provide
microsecond timing  accuracy. Currently, LXeGRIT events have a time stamp
derived from the microprocessor clock. This clock, however, does not have the
required stability and precision for pulsar timing analysis.

\begin{figure}[htb]
\bigskip
\bigskip
\centering
\includegraphics[bb=80 415 519 740,width=0.45\textwidth,clip]{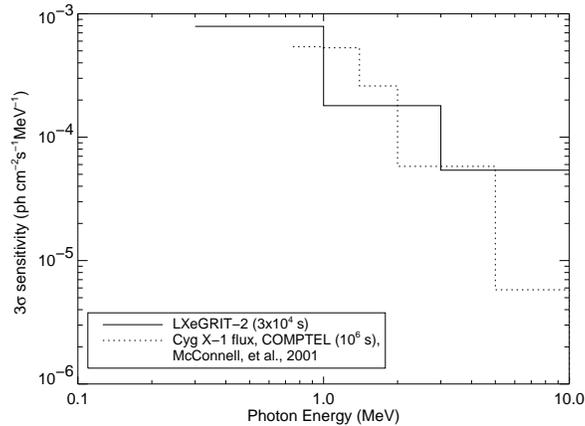}
\caption{LXeGRIT-2 continuum sensitivity}
\label{f:sensitivity}
\end{figure}

The combination of upgrades is expected to increase the detection efficiency for
multiple-site imaging events in the 1-3~\MeV\ band to the 3.5\% level, as
originally simulated for the LXeTPC. With 14~cm$^{2}$ effective area and the
background level measured in the last flight, reduced by about 15\% through
Compton imaging, we expect $\sim$730 source counts from the Crab and $\sim$1900
background counts in the the 1-3~\MeV\ band. This corresponds to a 3~$\sigma$
sensitivity of 3.7~$\times$ 10$^{-4}$ ph/cm$^{2}$/s, for a 3 $\times$ 10$^{4}$ s
balloon observation, or a 14 $\sigma$ Crab detection. 
With the improved LXeGRIT we will be able to look for polarization in the Crab
Nebula and pulsed spectra at \MeV\ energies, and could make sensitive
observations of other sources such as Cyg X-1, 3C273, GRS1915+105 and the
Orion Nebula.

\subsection{Towards a Next-Generation LXe-Based Compton Telescope}
\label{sec:fl2003}  
The overall efficiency of a next generation Compton telescope based on an array
of independent LXeTPCs could be maximized along the following lines:
\begin{enumerate}
\item Use thicker LXeTPC modules, with two or more drift regions for
improved photopeak efficiency and optimal electric field strength. Maximize the
information from the scintillation light, to effectively use this channel not
just for triggering, but for improved energy measurement and for background
rejection based on pulse shape discrimination. 
\item Develop a new DAQ system with fast DSP's to perform full online
event reconstruction and minimize the event rate to be transmitted to ground.
\item Develop a charge readout with improved spatial resolution,
replacing the current set of anodes, wires, and wire mesh. This would reduce 
the current requirements for matching X- and Y-wire signals as well as
wire/anode signals, which lead to significant event losses in the current
design. 
\end{enumerate}
This is a just short list of some improvements that would require more extensive
studies and development and are more suitable for a new detector module rather
than an upgrade of the current LXeGRIT.

\section*{Conclusions}

In this paper we have presented various laboratory and in-flight performance
results achieved in the period 1999 -- 2002 by the LXeGRIT collaboration. These
results show that the instrument is a fully tested, fully operational balloon
borne Compton telescope. The experience with the LXeGRIT prototype has also been
valuable to identify weaknesses of the current TPC design and signal readout,
for which we have presented solutions for an upgraded detector or a future new
detector module. 
LXeGRIT had a successful 27 hour balloon flight in the year 2000 and the
continuing analysis of the 40~GB of data collected at balloon altitude is giving
us a complete understanding of the performance expected from the instrument. The
measured \g-ray background in LXeGRIT at float altitude is in good agreement
with that expected from the known atmospheric and cosmic diffuse components
without having to invoke large contributions due to activation of Xe by
atmospheric neutrons or spallation-induced backgrounds in passive materials at
balloon altitude. Analysis improvements are currently underway, and we are
working at publishing a more in-depth report on the efficiency and in-flight
sensitivity of LXeGRIT in its 2000 configuration.
We are now planning to increase the sensitivity to astrophysical interesting
\g-ray sources. A new balloon flight with such improved detector would validate
this performance and help achieve the full science potential of LXeGRIT. 
Further studies are required for a next-generation LXe-based telescope.

\acknowledgments  
This work was supported by NASA under grant NAG5-5108.

\small
\bibliography{SR}  
\bibliographystyle{spiebib}   


\end{document}